\documentclass[lettersize,journal]{IEEEtran}
\usepackage{graphicx}
\usepackage{amsmath}
\usepackage{amssymb}
\usepackage{hyperref}
\usepackage{cite}
\usepackage{caption}
\usepackage[caption=false,font=normalsize,labelfont=sf,textfont=sf]{subfig}
\usepackage{subcaption}

\begin{document}
\title{Layered Architecture for Mobile Intelligence}
\author{Qingwen Liu and Mingqing Liu
}

\maketitle
\begin{abstract}
Artificial intelligence (AI) is rapidly evolving from a centralized computing capability into a pervasive infrastructure that interacts directly with the physical world. While recent perspectives highlight the roles of energy, chips, infrastructure, models, and applications in enabling large-scale AI systems, these frameworks primarily assume a static, cloud-centric computing paradigm. However, emerging intelligent applications, including autonomous vehicles, drones, robots, and wearable systems, require AI to operate in highly dynamic and mobile environments. This shift introduces mobility as a fundamental constraint across the entire AI ecosystem, affecting energy supply, computation, and intelligence deployment. In this article, we introduce the concept of the Mobile AI Stack, a mobility-aware architectural framework that integrates five tightly coupled layers: mobile energy networks, energy-efficient AI chips, cloud–edge–mobile infrastructure, distributed AI models, and embodied AI applications. The proposed framework provides a systematic perspective for understanding how energy delivery, computing architectures, communication networks, and AI algorithms must co-evolve to support large-scale mobile intelligence. We further discuss key research challenges and future directions toward building scalable, reliable, and energy-efficient mobile AI systems. Mobile AI Stack offers a conceptual blueprint of the next-generation infrastructure which deeply integrates the networks of computation, energy, and communications for mobile intelligence.
\end{abstract}
\begin{IEEEkeywords}
    Mobile Artificial Intelligence, AI Infrastructure, Wireless Energy Transfer, Embodied Intelligence
\end{IEEEkeywords}

\section{Introduction}

Artificial intelligence (AI) is rapidly evolving from a cloud-centric computing capability into a pervasive infrastructure that interacts directly with the physical world~\cite{ref1duan2023distributed}. Recent perspectives have described AI systems as a multi-layer technology stack spanning energy supply, computing hardware, digital infrastructure, AI models, and applications~\cite{ref2duan2023fl6g}. Such a view highlights that large-scale intelligence production depends not only on algorithms, but also on the underlying energy, computing, and networking ecosystems that sustain it~\cite{huang2026ai_cake,ref3wang2022edge_ai}. However, these architectural perspectives largely assume a static computing paradigm, where energy sources, computing infrastructure, and AI services remain geographically fixed.

At the same time, mobile communication networks are undergoing a profound transformation~\cite{ref4zhang2022edge_learning}. From the early days of providing ubiquitous connectivity to today’s broadband-centric systems, mobility has traditionally been treated as a secondary concern, primarily addressed through handover management, resource allocation, and link adaptation. In the emerging AI era, however, mobile networks are no longer merely conduits for data transmission. Instead, they are becoming platforms for intelligence creation, distribution, and execution, supporting a wide range of AI-enabled services that operate directly in dynamic physical environments~\cite{ref5zhou2023green_ai}. The rapid emergence of applications such as autonomous driving, immersive extended reality (XR), collaborative robotics, and large-scale Internet of Things (IoT) systems further amplifies this shift~\cite{ref6li2021uav_energy}. These applications require not only reliable communication but also continuous access to energy, real-time computation, and adaptive AI models. In many scenarios, intelligent devices must sense, learn, and act while moving through complex environments with limited energy supply and time-varying network conditions~\cite{ref7zeng2021uav_survey}. Consequently, the traditional assumption of static infrastructure becomes increasingly inadequate, as disruptions or constraints at the lower layers, such as energy availability or computing location, can directly affect higher-layer intelligence and application performance.

In this context, mobility extends far beyond user terminals. Energy itself can become mobile through wireless power transfer, energy harvesting, or energy sources mounted on vehicles and unmanned aerial platforms~\cite{ref8nguyen2021federated}. Computation can be mobile through edge servers, mobile base stations, and on-device AI accelerators~\cite{ref5zhou2023green_ai}. Moreover, chip evolution is from performance-centric to energy-efficient to support mobile applications~\cite{ref9tran2022split}. AI models can migrate across the network through mechanisms such as federated learning, split inference, and model adaptation~\cite{ref10park2023mobile_ai}. Ultimately, even decision-making intelligence may be distributed among multiple cooperative agents that dynamically learn and act in real time. This expanded notion of mobility fundamentally changes the design space of future networks by tightly coupling energy, communication, computation, and learning.

\begin{figure}
    \centering
    \includegraphics[width=1\linewidth]{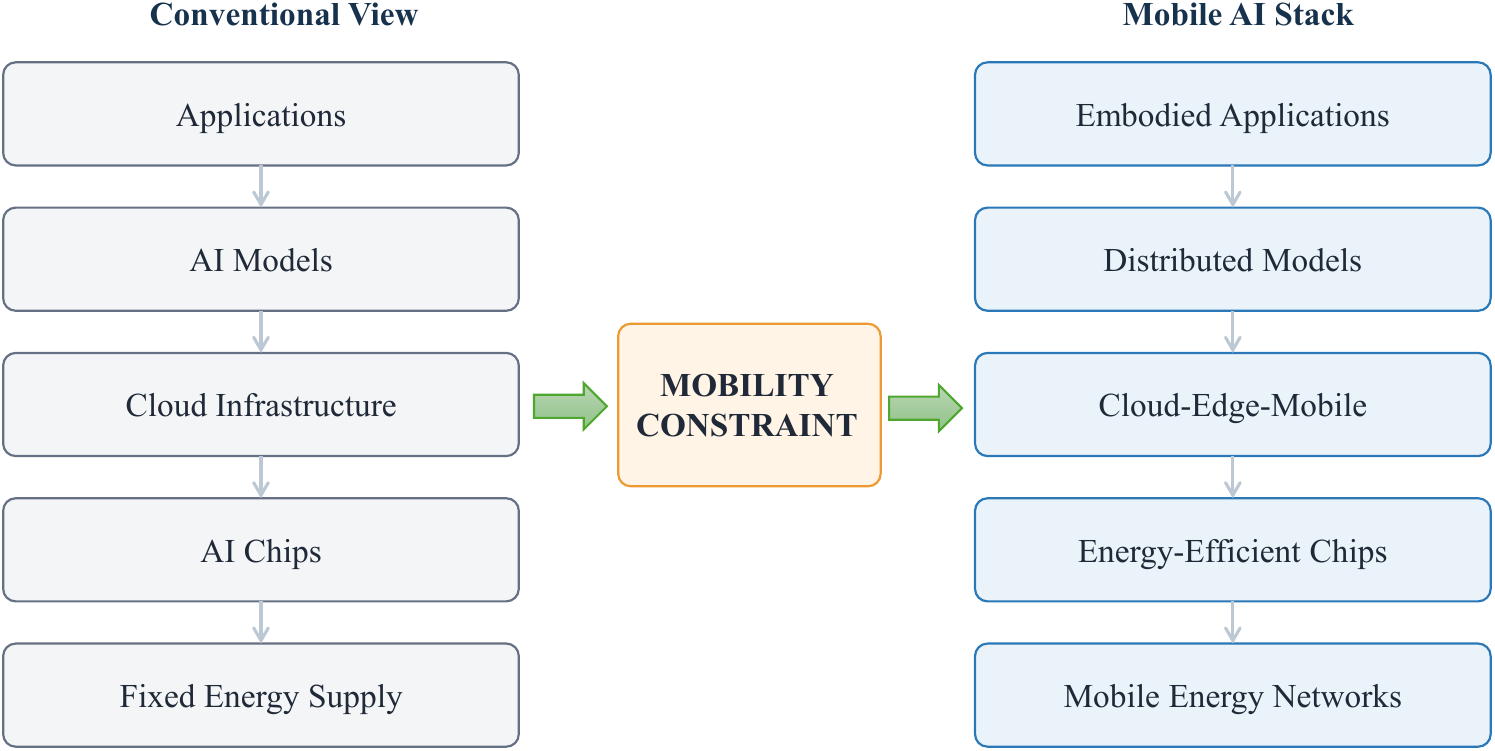}
    \caption{Origination of mobile AI stack.}
    \label{fig:evolution}
\end{figure}

Motivated by these trends, this article advocates a new paradigm referred to as mobile intelligence, in which intelligence generation, distribution, and execution are inherently mobility-aware. As in Fig.~\ref{fig:evolution}, rather than treating mobility as a challenge to be mitigated, mobile intelligence treats it as a first-class design principle shaping how energy, computation, models, and applications interact across the system. To systematically capture this paradigm, we introduce a mobility-centric five-layer architecture, referred to as the Mobile AI Stack. The proposed framework integrates mobile energy networks, energy-efficient AI hardware, cloud–edge–mobile infrastructure, distributed AI models, and embodied intelligent applications into a unified architectural perspective. By explicitly incorporating mobility into each layer of the AI ecosystem, the Mobile AI Stack provides a conceptual foundation for designing next-generation intelligent networks capable of supporting large-scale mobile AI services.

The main contributions of this article are summarized as:
\begin{itemize}
    \item We elevate mobility from a link-level or protocol-level concern to a system-wide design principle for AI-enabled mobile networks. Building upon this, we introduce the concept of mobile intelligence and articulate its key characteristics in terms of energy awareness, distributed computation, and mobility-adaptive learning.
    \item We propose a mobility-centric five-layer architecture that captures the interplay among energy, communication, computation, AI models, and applications.
    Then, we identify key enabling technologies and open research challenges that shape future mobile intelligence networks.
\end{itemize}

The remainder of this article is organized as follows. Section II discusses the background and motivation for mobility-aware AI infrastructure. Section III presents the mobility-centric five-layer architecture. Section IV  outlines open challenges and future research directions, and Section V concludes the article.

\section{Why Mobility Matters for AI Infrastructure}
Artificial intelligence is undergoing a fundamental transition from cloud-centric intelligence to ubiquitous and embodied intelligence. Early generations of AI systems were primarily deployed in centralized computing environments where large models were trained and executed in powerful data centers. In such architectures, data was collected from distributed devices, transmitted to the cloud, and processed by centralized algorithms. However, this paradigm is increasingly insufficient for emerging intelligent systems that operate directly in the physical world. Applications such as autonomous vehicles, robotic systems, drones, augmented reality devices, and mobile sensing platforms require real-time perception, reasoning, and decision-making within highly dynamic environments. These systems must interact continuously with the physical world while operating under strict constraints on latency, energy consumption, and connectivity.

Mobility introduces a set of fundamental challenges to traditional AI infrastructures, spanning energy, computation, communication, and their integration, as detailed below
\begin{itemize}
    \item First, energy availability becomes a critical constraint for mobile intelligent devices. Unlike data center servers that can rely on stable power supplies, mobile platforms are typically powered by limited onboard batteries, which restrict operational duration and computational capability. As AI workloads grow increasingly complex, ensuring sustainable energy support for mobile intelligence becomes a major challenge.
    \item Second, computation must be distributed across multiple layers of the network. Transmitting all sensory data to centralized cloud servers for processing is often impractical due to latency, bandwidth limitations, and reliability concerns. Instead, intelligent processing must occur across a hierarchy of computing resources, including cloud platforms, edge servers, and mobile devices.
    \item Third, communication infrastructure must support continuous connectivity for moving entities. Mobile AI systems frequently operate across heterogeneous networks and dynamically changing environments, requiring robust communication mechanisms capable of maintaining low-latency and high-reliability connections.
    \item Finally, mobility fundamentally changes the interaction between AI systems and the physical world. Instead of simply analyzing data, AI systems increasingly act through physical embodiments, including robots, vehicles, and autonomous machines. These systems must tightly integrate sensing, computing, communication, and actuation in order to function safely and effectively.
\end{itemize}

These challenges indicate that mobility is not an application-level feature but a cross-layer constraint that affects every component of the AI ecosystem. Supporting large-scale mobile intelligence therefore requires a new architectural perspective that explicitly incorporates mobility into AI infrastructure design. To address this need, we introduce the Mobile AI Stack, a layered framework that integrates mobile energy supply, efficient computing hardware, distributed infrastructure, collaborative models, and embodied applications into a unified architecture for next-generation intelligent systems.

\begin{figure}
    \centering
    \includegraphics[width=1\linewidth]{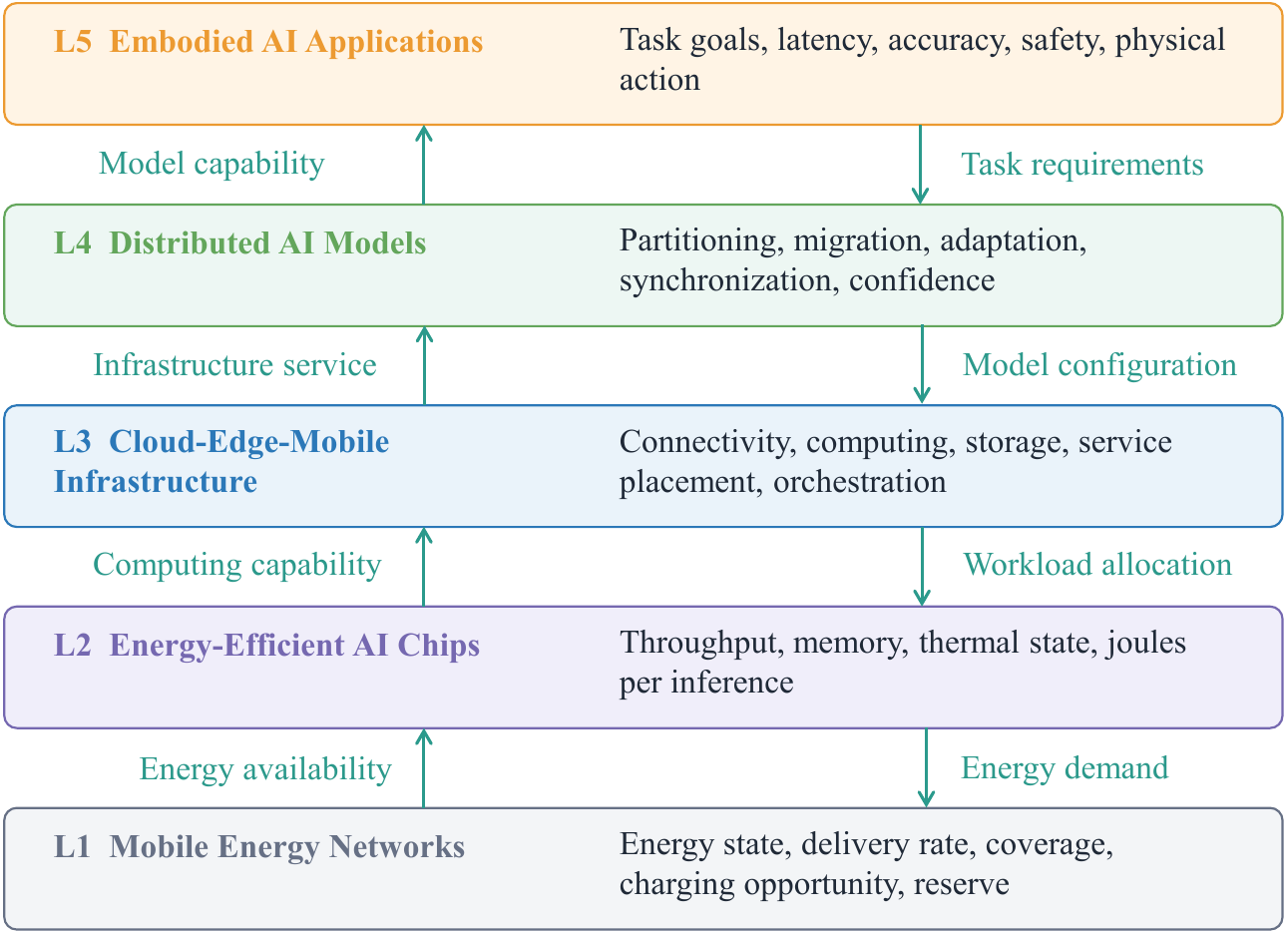}
    \caption{Mobility-centric five-layer intelligence architecture.}
    \label{fig:5layer}
\end{figure}

\section{Mobility-Centric Five-Layer Architecture}
The next generation of AI systems will increasingly operate in mobile and embodied environments, which require not only scalable intelligence but also mobility-aware infrastructure capable of supporting energy delivery, distributed computing, and real-time connectivity. We introduce the Mobile AI Stack, a layered architecture that integrates mobility considerations across the entire AI ecosystem. As illustrated in Fig. \ref{fig:5layer}, the proposed stack consists of five tightly coupled layers: mobile energy networks, energy-efficient AI chips, cloud–edge–mobile infrastructure, distributed AI models, and embodied applications. Each layer provides critical capabilities that collectively enable scalable, mobile, and real-time intelligent systems.

\subsection{Mobile Energy Networks}

\begin{figure}[t]
    \centering
    \includegraphics[width=1\linewidth]{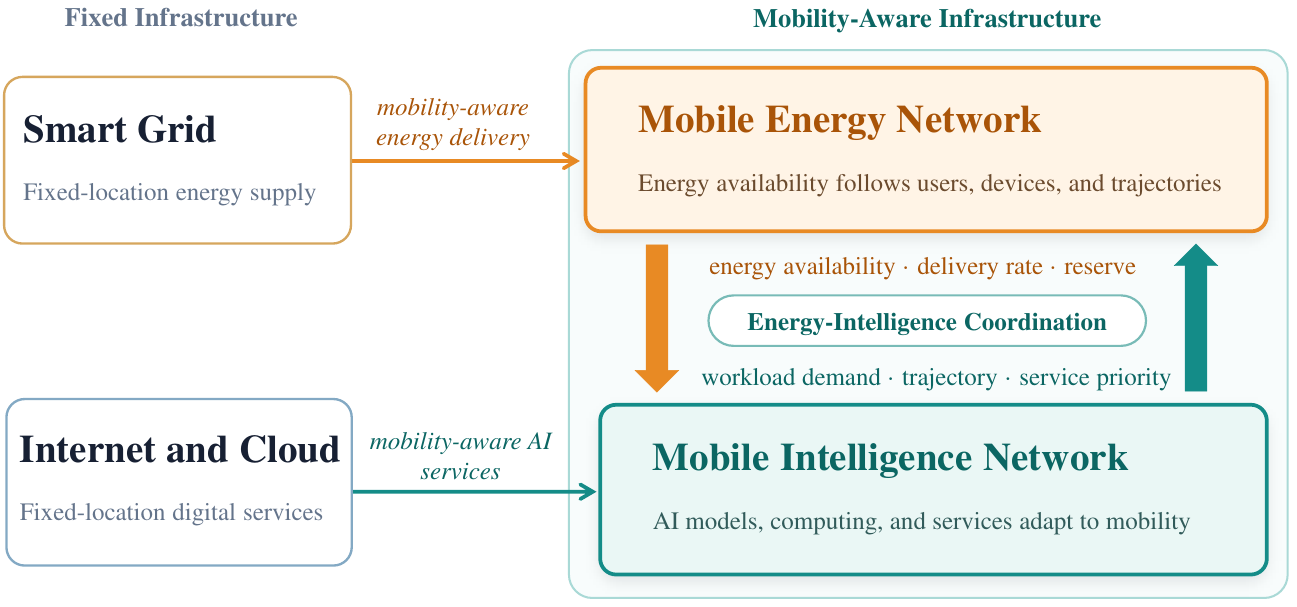}
    \caption{Positioning of mobile intelligence network.}
    \label{fig:posi}
\end{figure}

At the foundation of the Mobile AI Stack lies the mobile energy layer, which ensures that intelligent devices operating in dynamic environments have continuous and reliable access to power. Traditional AI infrastructure relies primarily on centralized energy supply within large-scale data centers. In contrast, mobile AI systems, e.g.,  autonomous robots, drones, and edge sensing platforms, operate in distributed environments where continuous wired power supply is often infeasible. Consequently, the availability and delivery of energy become fundamental constraints for mobile intelligence. Mobile energy networks aim to address this challenge by enabling flexible, spatially distributed energy provisioning for mobile AI systems as shown in Fig.~\ref{fig:posi}. Potential technologies include wireless power transfer, laser-based energy delivery, energy harvesting, and distributed charging infrastructures. These mechanisms allow energy to be delivered dynamically to mobile devices, thereby extending operational duration and enabling persistent intelligent services.
By extending energy supply from fixed physical connections, mobile energy networks significantly expand the operational space of AI systems. This capability is particularly critical for applications such as autonomous drones, mobile sensing, and large-scale robotic systems, where continuous operation and wide-area mobility are essential.

\subsection{Energy-Efficient AI Chips}

Above the energy layer lies the energy-efficient AI chip layer, which is responsible for transforming available energy into computational intelligence. While modern AI training primarily occurs in large data centers equipped with high-performance accelerators, mobile AI applications impose fundamentally different constraints. Devices operating in mobile environments must perform inference with strict limits on power consumption, thermal dissipation, and physical size. As a result, the design objective of mobile AI processors shifts from maximizing absolute computational throughput to maximizing computational efficiency per unit energy. Emerging technologies in this layer include edge AI accelerators, heterogeneous computing architectures, neuromorphic processors, and specialized inference chips optimized for deep learning workloads. These processors aim to deliver high inference performance while maintaining extremely low energy consumption. Energy-efficient AI chips will enable intelligent capabilities to migrate from centralized cloud platforms toward edge devices. This shift significantly reduces latency, improves reliability, and allows intelligent systems to operate even in environments with limited network connectivity.

\begin{figure}
    \centering
    \includegraphics[width=1\linewidth]{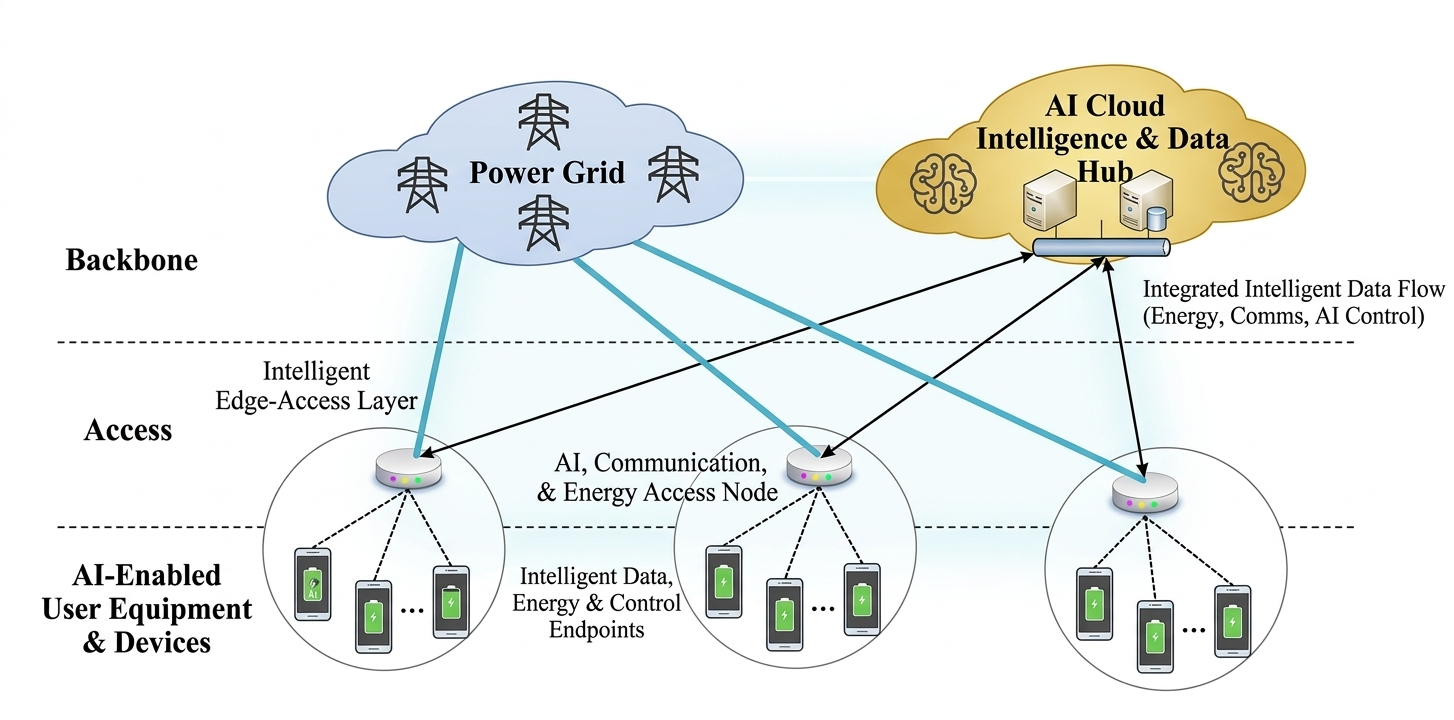}
    \caption{Mobile intelligence network architecture.}
    \label{fig:flow}
\end{figure}
\subsection{Cloud–Edge–Mobile Infrastructure}

As in Fig.~\ref{fig:flow}, the third layer of the Mobile AI Stack is the cloud–edge–mobile infrastructure, which provides the distributed computing and communication backbone required to support large-scale mobile intelligence. Traditional AI systems typically rely on centralized cloud infrastructure for both training and inference. However, as intelligent applications become increasingly mobile and latency-sensitive, purely cloud-based architectures become insufficient. Real-time perception, decision-making, and control require computing resources that are physically closer to the data sources and actuators. To address this challenge, the Mobile AI Stack adopts a hierarchical computing architecture consisting of cloud data centers, edge computing nodes, and mobile devices. In this architecture, the cloud provides large-scale model training and global coordination, edge nodes support low-latency inference and regional services, and mobile devices perform local perception and control. This distributed infrastructure enables efficient allocation of computational resources across different layers of the network. It also facilitates scalable deployment of intelligent services while maintaining responsiveness and reliability in highly dynamic environments.

\subsection{Distributed AI Models}
Built upon the distributed infrastructure layer are distributed AI models, which enable collaborative intelligence across cloud, edge, and device platforms. Traditional AI models are often designed to operate as monolithic systems deployed on centralized servers. However, mobile AI applications require models that can operate across heterogeneous hardware platforms and dynamically adapt to varying computational and communication constraints. Distributed AI models address this challenge by partitioning learning and inference processes across multiple computational nodes. Techniques such as split learning, federated learning, collaborative inference, and model compression enable different components of an AI system to execute across cloud, edge, and mobile devices. This distributed modeling paradigm offers several advantages. First, it reduces communication overhead by performing local processing near the data source. Second, it enhances privacy and security by keeping sensitive data on local devices. Third, it improves scalability by enabling large populations of devices to collaboratively contribute to the learning process.

\subsection{Embodied AI Applications}
At the top of the Mobile AI Stack are embodied AI applications, where artificial intelligence interacts directly with the physical world. Unlike traditional software applications that operate purely within digital environments, embodied AI systems are integrated into physical entities such as autonomous vehicles, robots, drones, wearable devices, and smart infrastructure. These systems must continuously perceive their surroundings, reason about complex environments, and execute actions in real time. Embodied AI applications therefore impose stringent requirements on the entire AI stack, including low-latency computation, reliable connectivity, efficient energy usage, and robust sensing capabilities. The effectiveness of these applications depends critically on the coordinated operation of all underlying layers in the Mobile AI Stack. Examples of embodied AI include autonomous transportation systems, industrial robotics, intelligent logistics platforms, and immersive augmented reality environments. These applications represent some of the most transformative opportunities for AI, as they extend intelligence beyond digital systems and into the physical world.

\section{Challenges and Open Research Directions}
While the Mobile AI Stack provides a conceptual framework for enabling large-scale mobile intelligence, realizing this vision requires significant advances across multiple technological domains. In this section, we outline several key research challenges that must be addressed to support the next generation of mobility-aware AI systems.

\subsection{Mobile Energy Delivery}
Sustaining intelligent operations in mobile environments requires new mechanisms for energy provisioning. Conventional battery-based solutions impose strict limitations on operating time, particularly for power-hungry AI workloads such as real-time perception and autonomous decision-making.
Future research should explore mobile energy networks capable of dynamically supplying energy to distributed devices. Technologies such as wireless power transfer, laser-based energy delivery, and energy harvesting systems offer promising solutions for enabling persistent operation of mobile AI platforms. Integrating these energy delivery mechanisms with communication and control networks represents an important research direction for large-scale mobile AI ecosystems.

\subsection{Ultra-Efficient Edge AI Hardware}
Mobile AI devices must perform increasingly complex inference tasks while operating under strict power and thermal constraints. Designing hardware that can deliver high computational performance within extremely limited energy budgets remains a major challenge. Emerging research directions include specialized AI accelerators optimized for edge inference, heterogeneous computing architectures that combine multiple processing units, and neuromorphic computing paradigms that mimic the efficiency of biological neural systems. Future processors must not only improve computational efficiency but also support adaptive workload distribution across cloud, edge, and device platforms.
\subsection{Scalability Under Mobility}

Scalability remains a primary challenge as mobile intelligence systems expand to large numbers of highly mobile agents, such as connected vehicles, UAV swarms, and dense IoT deployments.
Frequent topology changes, intermittent connectivity, and heterogeneous mobility patterns can severely degrade the performance of centralized coordination and learning mechanisms.
Existing solutions often struggle to maintain stable performance as the scale and mobility intensity increase.
Open research directions include mobility-aware hierarchical architectures, decentralized control mechanisms, and learning frameworks that gracefully degrade under partial observability.
Understanding how collective intelligence emerges and scales under massive mobility, while maintaining robustness and efficiency, remains a largely unexplored theoretical problem.

\subsection{Model Migration, Consistency, and Lifelong Learning}
Mobile intelligence relies on the seamless migration of models, tasks, and learned knowledge across devices, edge servers, and cloud platforms.
However, frequent migration raises challenges in model consistency, version control, and knowledge alignment, especially when learning occurs concurrently at multiple locations.
Ensuring coherent intelligence evolution under mobility requires new mechanisms for distributed model synchronization, conflict resolution, and continual learning.
Open directions include lightweight model versioning, consistency-aware federated learning, and lifelong learning frameworks that prevent catastrophic forgetting while adapting to evolving environments.

\subsection{Security, Trust, and Privacy in Mobile Intelligence}
The distributed and mobile nature of intelligent systems significantly expands the attack surface.
Mobile agents may be vulnerable to data poisoning, model manipulation, spoofing, and adversarial interference, while sensitive sensory data raises serious privacy concerns.
Future research must integrate security and trust as first-class design objectives rather than afterthoughts.
This includes secure learning protocols, trust-aware collaboration mechanisms, and privacy-preserving intelligence sharing.
For embodied intelligence, additional safety-critical concerns arise, as compromised decisions may directly translate into physical harm, making robust and verifiable intelligence an urgent requirement.

\subsection{Standardization and Ecosystem Integration}
Finally, the realization of mobile intelligence at scale requires coherent standardization across communication, computation, energy, and intelligence interfaces.
Current systems are often developed in isolated vertical silos, hindering interoperability and large-scale deployment.
Open research directions include unified abstractions, cross-domain interfaces, and standard benchmarks that span multiple layers of the mobile intelligence stack.
Building an open ecosystem that aligns network operators, device manufacturers, AI developers, and application providers remains a critical challenge.

\section{Conclusion}
In this article, we introduced the concept of the Mobile AI Stack, a mobility-aware architectural framework that integrates mobile energy networks, energy-efficient AI chips, cloud–edge–mobile infrastructure, distributed AI models, and embodied applications. By explicitly incorporating mobility considerations into each layer of the AI ecosystem, the proposed framework provides a systematic perspective for understanding and designing next-generation intelligent systems. The transition toward mobile AI will reshape not only computing architectures but also energy networks, communication systems, and industrial ecosystems. Addressing the associated challenges requires interdisciplinary collaboration across artificial intelligence, wireless communications, energy systems, and robotics. As these technologies continue to converge, the Mobile AI Stack may serve as a foundational blueprint for the future of large-scale intelligent infrastructure.

\bibliographystyle{IEEEtran}
\bibliography{reference}
\end{document}